\documentclass[pre,amssymb,floats,twocolumn,showpacs,superscriptaddress]{revtex4}

\usepackage{epsfig}

\begin{document}

\title{Self-organization of collaboration networks}

\author{Jos{\'e} J. Ramasco}
\email{jjramasc@fc.up.pt}
\affiliation{Departamento de F{\'\i}sica and Centro de F{\'\i}sica do Porto,
Faculdade de Ci{\^e}ncias, Universidade do Porto, Rua do Campo Alegre
687, 4169-007 Porto, Portugal.}

\author{S. N. Dorogovtsev}
\affiliation{Departamento de F{\'\i}sica, Universidade de Aveiro, Campus 
Universitario de Santiago, 3810-193 Aveiro, Portugal.}
\affiliation{Laboratory of Physics Helsinki University of 
Technology FIN-02015 HUT, Finland.}
\affiliation{A. F. Ioffe Physico-Technical Institute, 194021
  St. Petersburg, Russia.}

\author{Romualdo Pastor-Satorras}
\affiliation{Departament de F{\'\i}sica i Enginyeria Nuclear, Universitat
  Polit{\`e}cnica de Catalunya, Campus Nord, 08034 Barcelona, Spain.}

\date{\today}

\begin{abstract}
  We study collaboration networks in terms of evolving,
  self-organizing bipartite graph models.  We propose a model of a
  growing network, which combines preferential edge attachment with
  the bipartite structure, generic for collaboration networks. The
  model depends exclusively on basic properties of the network, such
  as the total number of collaborators and acts of collaboration, the
  mean size of collaborations, etc. The simplest model defined within
  this framework already allows us to describe many of the main
  topological characteristics (degree distribution, clustering
  coefficient, etc.) of one-mode projections of several real
  collaboration networks, without parameter fitting.  We explain the
  observed dependence of the local clustering on degree and the
  degree--degree correlations in terms of the ``aging'' of
  collaborators and their physical impossibility to participate in an
  unlimited number of collaborations.

\end{abstract}


\maketitle

\section{Introduction}\label{s-introduction} 

Recent years have witnessed an upsurge in the study of complex
systems that can be described in terms of networks, in which the
vertices picture the elementary units composing the system, and the
edges represent the interactions or relations between pairs of units
\cite{barabasi02,dm03}. These studies have led to the
development of a modern theory of complex networks 
which 
has found 
fruitful
applications in fields as diverse as the Internet
\cite{romuvespibook}, the World-Wide Web \cite{www99}, or biological
interacting networks \cite{wagner01,jeong01,spsk,vazquez}.

An important example of this kind of systems, that has attracted a
great deal of interest from researchers in different scientific
fields, are social networks \cite{wass94}. The study of social
networks has been traditionally hindered by the small size of the
networks considered and the difficulties in the process of data
collection (usually from questionnaires or interviews). More recently,
however, the increasing availability of large digital databases has
allowed to study a particular class of social networks, the so-called
\textit{collaboration networks}. These networks can be defined in a
non-ambiguous way, and their exceptionally large size has permitted
empirical researchers to obtain a reliable statistical description of
their topological properties and to arrive at solid conclusions
concerning their structure.

Social collaboration networks are generally defined in terms of a set
of people (called \textit{actors} in the social science literature),
and a set of \textit{collaboration acts}. Actors relate to each other
by the fact of having participated in a common collaboration act.
Examples of this kind of networks can be found in movie actors related
by co-starring the same movie, scientist related by co-authoring a
scientific paper, members of the boards of company directors related
by sitting on the same board, etc. Collaboration networks can be
represented as bipartite graphs \cite{chartrand} with two types of
vertices, one kind representing the actors, while vertices of the
other kind are acts of collaboration.  As a rule, however, it is the
one-mode projections of these bipartite graphs that are empirically
studied. In these projections, the vertices representing the acts of
collaboration are excluded, and collaborating pairs of actors are
connected by edges.  Since multiple connections in the projected graph
are usually ignored, the projection is less informative than the
original bipartite graph.

The study of several examples of large collaboration networks
\cite{albert00,watts98,newman01a,newman01b,schubert} 
allows one 
to
draw a number of conclusions regarding the main topological properties
of 
one-mode projections of these networks:
\begin{enumerate}
\item The degree distribution $P(k)$, defined as the probability that
  a vertex is connected to $k$ other vertices, 
often 
exhibits a fat tail,
  that can be approximated by a power law behavior for large $k$.
\item The clustering coefficient, roughly defined as the probability
  that two neighbors of any given vertex are also neighbors of each
  other, takes in average large values, and it locally depends on the
  vertex degree, signaling the presence of a structure in the network 
  \cite{ravasz1,ravasz2}.
\item The degrees 
of the nearest neighbor vertices 
are positively correlated,  
i.e.,  
  vertices with large degree have a high probability to be connected
  to vertices with large degree, and vice-versa. This property has
  been dubbed \textit{assortative mixing} \cite{assortative}.
\end{enumerate}

The general presence of these three properties in most collaboration
graphs prompts toward the development of models capable to reproduce
and explain these features.  In general, the first insight into the
architecture of a complex network is provided by ``formal''
constructions of random graphs.  These constructions allow one to
reproduce the structure of complex networks, but completely ignore the
mechanisms underlying these architectures.  The minimal formal model
of a complex one-partite graph, that is a graph composed by a single
type of vertices, is the configuration model
\cite{bekessi72,benderoriginal,bollobas1980,wormald}.  In simple
terms, the configuration model generates (uncorrelated) graphs, which
are maximally random under the constraint that their degree
distribution is a given one.  Similarly, the minimal model of a
complex bipartite graph is a bipartite network that is maximally
random under the constraint that the two degree distributions for both
kinds of vertices are given \cite{newman01c,latapy}. One can see that
this is a direct generalization of the configuration model to
bipartite graphs.  The quality of the configuration model applied to
bipartite graphs was checked in Ref.~\cite{newman01c}.  In this work,
it was proved that the empirical degree distribution of the one-mode
projection of a bipartite collaboration graph agrees with that of the
configuration model when the empirically observed degree distributions
are imposed on the two kinds of vertices.  One should emphasize that a
one-mode projection of an uncorrelated bipartite graph is correlated.
In particular, this projection contains numerous triangles of edges
which results in a high clustering \cite{newmansocialdiff}.

Therefore, it might seem at first sight that, in order to explain the
nature of the structure of collaboration networks, it is sufficient
(i) to propose a mechanism generating the specific degree
distributions of the two kinds of vertices and afterwards (ii) to
connect vertices by using the configuration model. This approach,
however, fails to reproduce the complex distribution of connections
over collaboration networks, since assumes pure randomness.  Also, it
does not explain specific distributions of vertex degrees, which, in
the configuration model, are assumed to be given.  Note that, while
providing reasonable values of clustering, the configuration model
fails to reproduce the type of degree--degree correlations in
collaboration networks.  Consequently, in order to fully explain the
specific architecture of collaboration networks (fat-tailed degree
distributions, high clustering, assortative mixing, etc.), we have to
introduce a mechanism for the linking of vertices in these networks.

In the present paper we propose a first approximation to such a
mechanism.  In our approach, we treat collaboration networks as
growing, self-organizing, correlated bipartite graphs, applying the
ideas at the basis of the preferential attachment concept put forward
by Barab{\'a}si and Albert \cite{barab99} in the network modeling
context (see also Ref.~\cite{simon55}) to bipartite graphs. The
simplest model that we can define already allows us to quantitatively
describe most of the empirical data on collaboration networks without
fitting, only by using basic numbers characterizing the real networks.
We emphasize that the absence of fitting convincingly proves the
validity of the concept.

The degree--degree correlations in the one-mode projections of
collaboration graphs are a topic of our special interest. We show that
the ``assortative mixing'' character of these correlations are not so
inevitable in collaboration networks, as it is usually believed
\cite{newmansocialdiff}. We explain the origin of the assortative
mixing in real collaboration networks in terms of the aging of actors,
which cannot accept new connections during the whole growth process of
the network.

The present paper is organized as follows. In
Sec.~\ref{s-organization} we review 
measurements defined to
characterize the topological properties of collaboration 
networks---bipartite graphs and their one-mode projections.  
Sec.~\ref{sec:empir-data-coll} presents the
existing empirical data on collaboration networks, referring in
particular to the networks of movie actors, scientific coauthorship,
and company directors.  In Sec.~\ref{s-model} we introduce a simple
model of a growing, self-organizing bipartite graph.
Sec.~\ref{s-results} contains results obtained for this model and a
detailed comparison with empirical data. Separately, in
Sec.~\ref{s-model-age} we discuss and explain the presence of positive
correlations between the degrees of the nearest-neighbor vertices in
collaboration networks. In this Section we discuss the importance of
the ``physical'' limitation of vertex degrees in collaboration
networks. Finally, in Sec.~\ref{s-conclusions} we draw the main
conclusions of our work.

\section{Structural organization of collaboration
  networks}\label{s-organization}
 
As we have already mentioned in the Introduction, collaboration
networks can be represented as bipartite graphs \cite{note1}. 
On one side, we have
collaboration acts (e.g. movie co-starring or paper co-authorship,
belonging to the same company, school, etc.), that may be represented
as a special kind of vertices. On the other side we have the actors
(normal vertices), that are linked to the collaborations acts in which
they participate.  Two independent degree
distributions may then be defined: First, the probability $S(n)$ of
having $n$ actors participating in any collaboration act; and second,
the probability $Q(q)$ that any actor has taken part in $q$
collaboration acts.

In most of cases, however, the object of study is not the whole
bipartite graph but its one-mode projection: i.e., the network formed
by the collaborating actors linked to each other whenever they have
shared a collaboration act. For this projected
network, another degree distribution $P(k)$ may be considered, defined
as the probability that any given actor is connected to $k$ others.
Focusing on the one-mode projection of a collaboration network, many
other properties generally studied in common random graphs can 
be
measured. This type of study has already been carried out for several
empirical social networks (see Ref.~\cite{newman-review} for a recent
review). The quantities that we use to describe the structure of the
projected network are the clustering coefficient 
and the mean clustering \cite{watts98}, the
average clustering coefficient of vertices of degree $k$
\cite{alexei02,d01,s02,ravasz02}, the average degree of the nearest neighbors
of the vertices of degree $k$ \cite{alexei}, and the Pearson
correlation coefficient defined in
Refs.~\cite{assortative,newmanmixing}.

The local clustering $c_i$ of the vertex $i$ is given by the rate
between the number of triangles connected to that vertex, $s_i$, and
the total number of possible triangles including it, $k_i\,
(k_i-1)/2$, i.e.
\begin{equation}
  c_i=\frac{2 s_i}{k_i(k_i-1)}.
\label{eq:4}
\end{equation}
To obtain the mean degree-dependent local clustering we average the
local clustering over all vertices with degree $k$ in a network,
\begin{equation}
c(k)  =  \frac{s(k)}{k(k-1)/2} 
\, , 
\label{e1}
\end{equation}
where $s(k) = \langle s_i(k) \rangle$ is the mean number of connections 
between 
    the nearest neighbors of a vertex of degree $k$. 
The mean clustering $\langle c \rangle$ is defined as the average of the local
clustering over all the vertices in 
a 
network, i.e.
\begin{equation}
\langle c \rangle  = \sum_{k>1} P(k) \, c(k) = \frac{1}{N} \sum_i c_i,
\label{e2}
\end{equation}
where $N$ is the total number of actors (vertices), and the second sum
runs over the $N$ vertices of the network. 
The clustering coefficient of a graph (transitivity in sociology 
\cite{wass94})  
is 
defined as 
\begin{eqnarray}
c  
& = &  \frac{ 
3 \times \mbox{number of triangles of edges in 
a 
graph}
}
    {
\mbox{number of connected triples of vertices} 
}  \nonumber \\[5pt]
    & =& \frac{ 2 \, \sum_k P(k)\, s(k)}{\sum_k P(k) \, k \, (k-1)}
\, . 
    \label{e3} 
\end{eqnarray}
The quantities $c(k)$, $\langle c \rangle$ and 
$c$ 
provide 
information on the concentration of loops of length three in 
a 
graph, which
is typically 
high in social networks \cite{socialwebs}. 
Note that if the local clustering depends on the degree, $c \neq \langle c \rangle$, and the (relative) difference is great in many real-world networks.   

The correlations between the degrees of connected vertices can be
fully defined by means of the joint probability $P(k, k')$, defined
such that $(2-\delta_{k, k'}) P(k, k')$ is the probability that a randomly
chosen edge connects to vertices of degree $k$ and $k'$
\cite{marianproc}. 
($\delta_{k, k'}$ is the Kronecker symbol.) 
By using 
this quantity 
one 
can compute the average
degree of the nearest neighbors of the vertices of degree $k$,
$\bar{k}_{nn}(k)$, defined as
\begin{equation}
  \bar{k}_{nn}(k) = \langle k \rangle \frac{\sum_{k'} k' P (k, k')}{k P (k)} \equiv
  \sum_{k'} k' P( k' \vert k).
\end{equation}
where $P( k' \vert k)$ is the conventional probability that a vertex
of degree $k$ is connected to a vertex of degree $k'$.  In simple
terms, if the network presents assortative mixing (large degree
vertices connect preferably with large degree vertices, and
vice-versa), $\bar{k}_{nn}(k)$ increases with $k$ \cite{note4}.  In the
case of disassortative mixing (large degree vertices connected with
low degree vertices, and vice-versa), $\bar{k}_{nn}(k)$ is conversely
a decreasing function of $k$. Analogous information can be obtained by
means of the Pearson correlation coefficient, defined as
\begin{equation}
  r = \langle k \rangle \frac{\sum_kk^2 \bar{k}_{nn}(k) P(k) -  \langle k^2 \rangle^2}{ \langle k \rangle
    \langle k^3 \rangle -  \langle k^2 \rangle^2}.
\end{equation}
Here positive (negative) values of $r$ imply the presence of
assortative (disassortative) mixing.

\section{Empirical data on collaboration networks}
\label{sec:empir-data-coll}

\begin{figure}
\epsfxsize=80mm
\epsffile{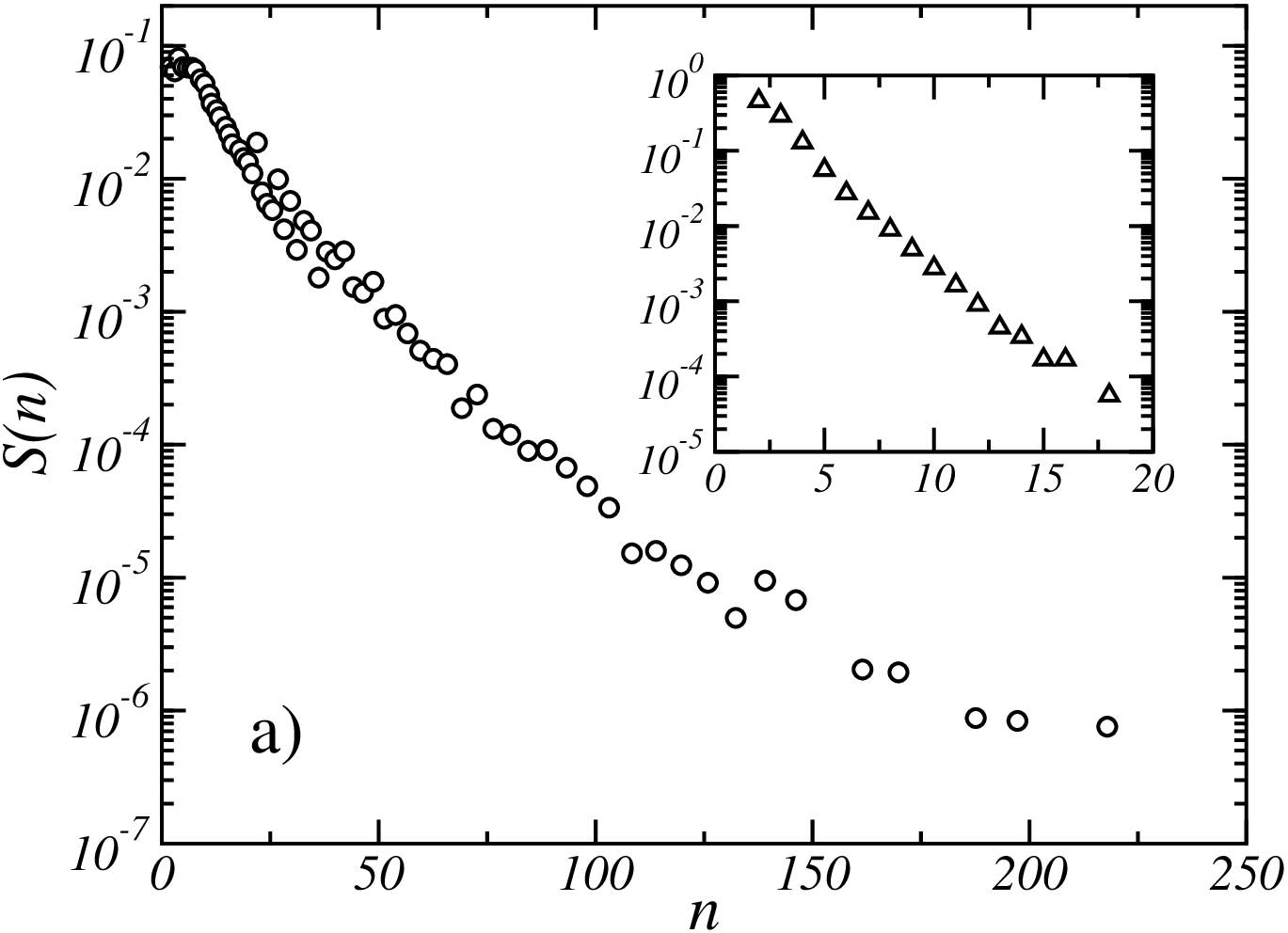}
\epsfxsize=80mm
\epsffile{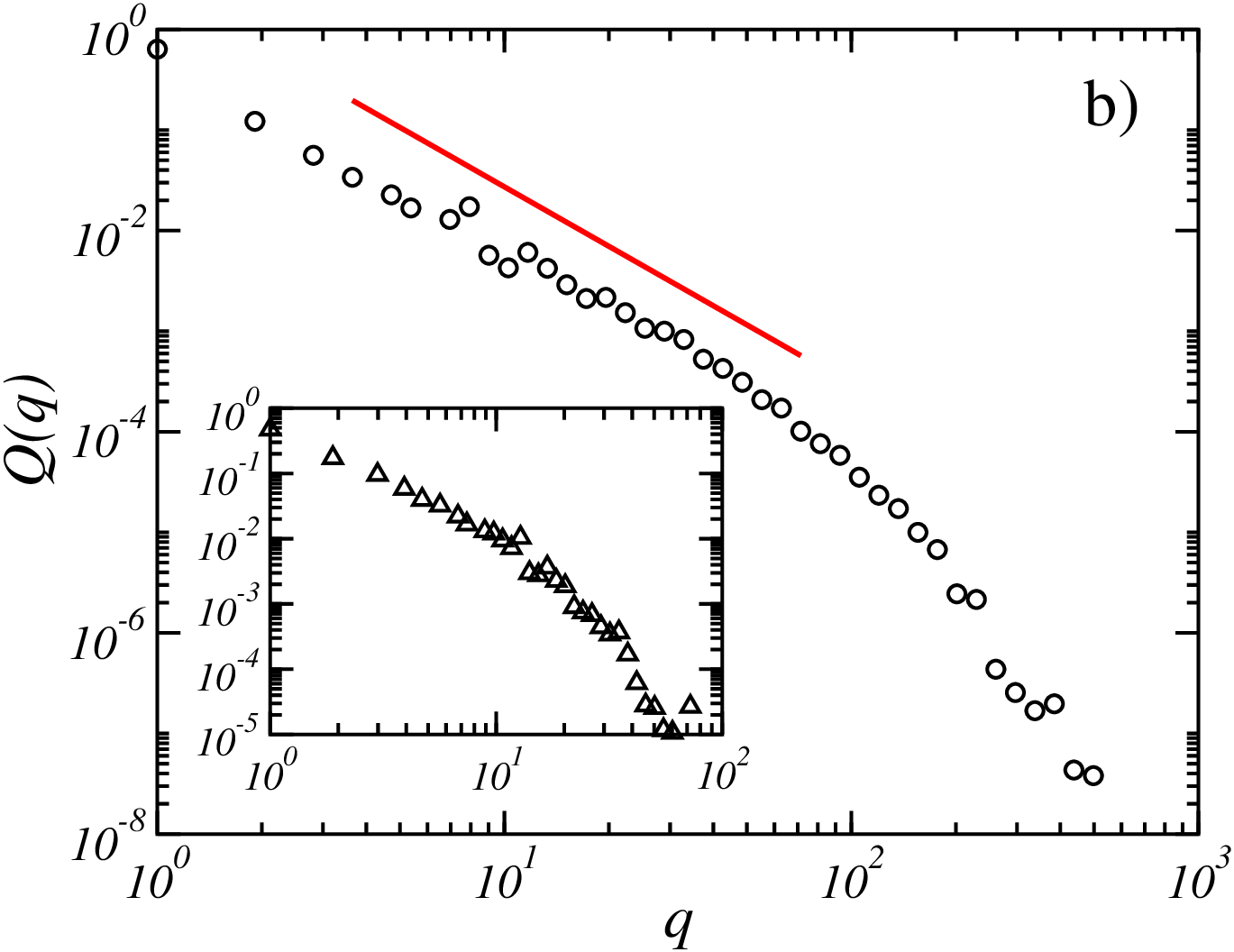}
\caption{a) Probability distribution of the size of collaboration acts
  $S(n)$ for the movie actors collaboration network (main plot) and
  the scientific collaboration network (inset). %
  b) Probability distribution that an actor has taken part in $q$
  collaboration acts $Q(q)$ for the movie actors collaboration network
  (main plot) and the scientific collaboration network (inset). The
  solid line has slope $\simeq 2$.}
\label{f1}
\end{figure}

In the present Section we revisit the empirical analysis of three
typical social collaboration networks. We consider in particular the
network formed by movie actors playing in the same movie, the network
of scientific collaborations, and the network of company board
directors sitting on the same board.

\subsection{Movie actor collaboration network}

The movie actor collaboration network that we consider was obtained
from the Internet Movie Database (IMDB) \cite{note2}.  Taking only
into account movies with more than one actor, and discarding
duplicated actors in several movies, we finally analyze the properties
of a network composed by $N = 382219$ actors acting on $t = 118477$
films. The distribution of movie cast size, $S(n)$, is represented in
Fig.~\ref{f1}a).  Apparently, this function follows an exponential
decay, with an average cast size of $\overline{n} = 12.33$ actors per
movie.  The distribution $Q(q)$ (number of movies in which an actor
has played) adjusts better to a power law decay $Q(q) \sim q^{- \gamma} $
with an apparent exponent $\gamma \approx 2$, see Fig.~\ref{f1}b).  An upper
cutoff of this dependence is observed around $q_{c} \sim 100$.  The mean
number of movies played per actor is $\langle q\rangle = 3.82$.

\begin{figure}
\epsfxsize=80mm
\epsffile{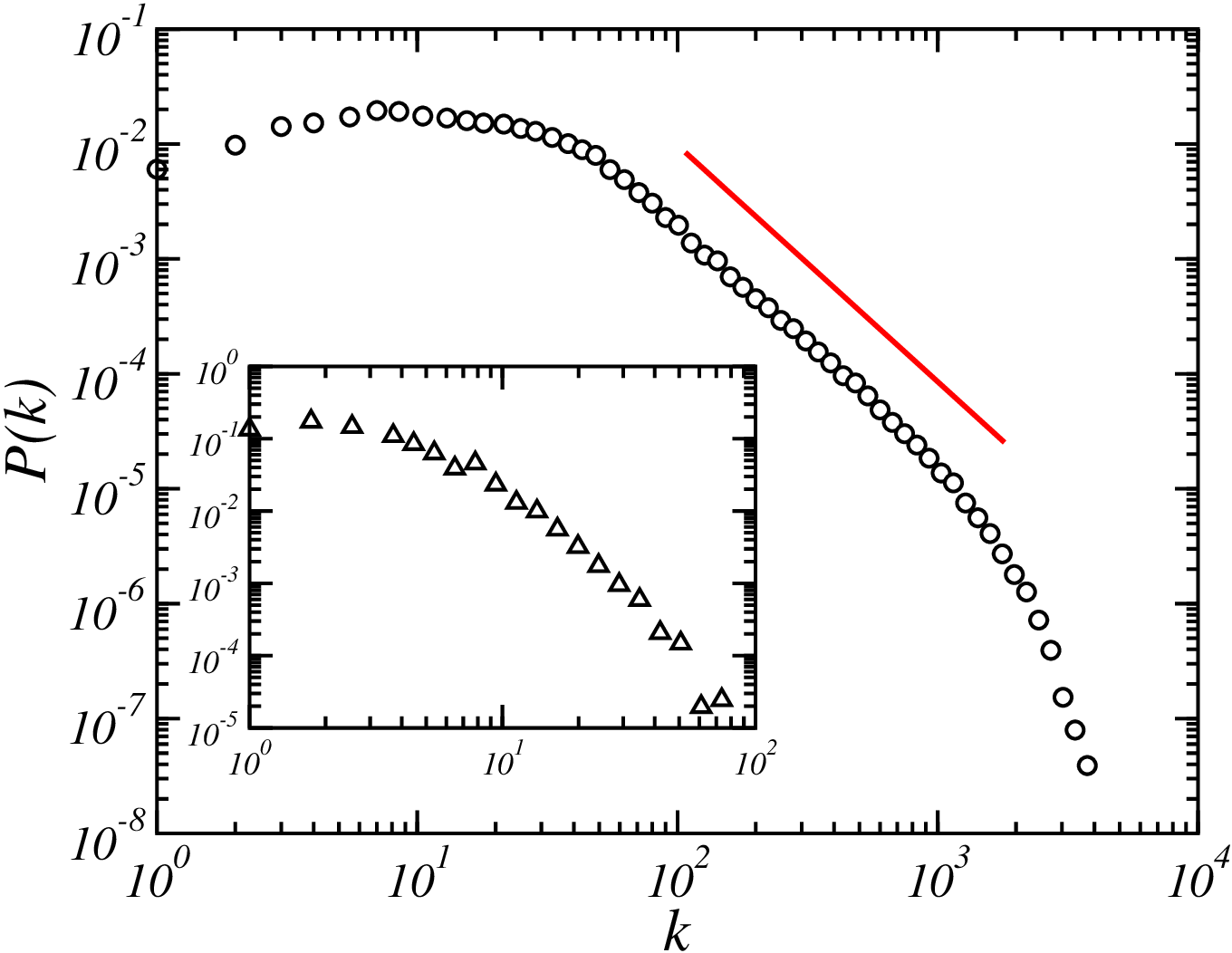}
\caption{Degree distribution $P(k)$ of the one-mode projection for
  the movie 
  actors collaboration network (main plot) and the scientific
  collaboration network (inset). The full line has slope $\simeq 2$.}
\label{fpk}
\end{figure}

\begin{figure}
\epsfxsize=80mm
\epsffile{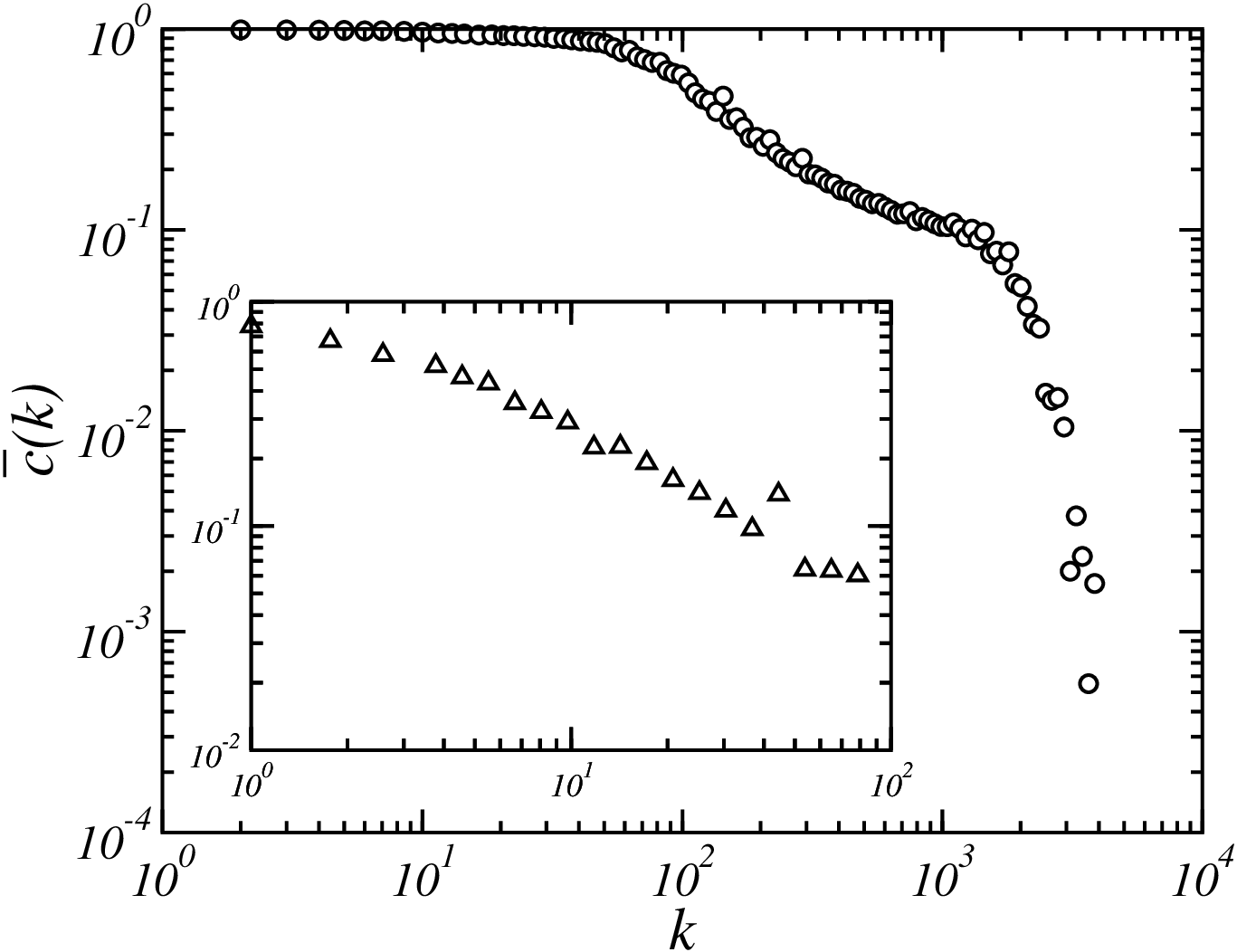}
\caption{
Local clustering as a function of the degree $c(k)$ for
  the movie actors collaboration network (main plot) and the
  scientific collaboration network (inset).}
\label{figck}
\end{figure}

\begin{figure}
\epsfxsize=80mm
\epsffile{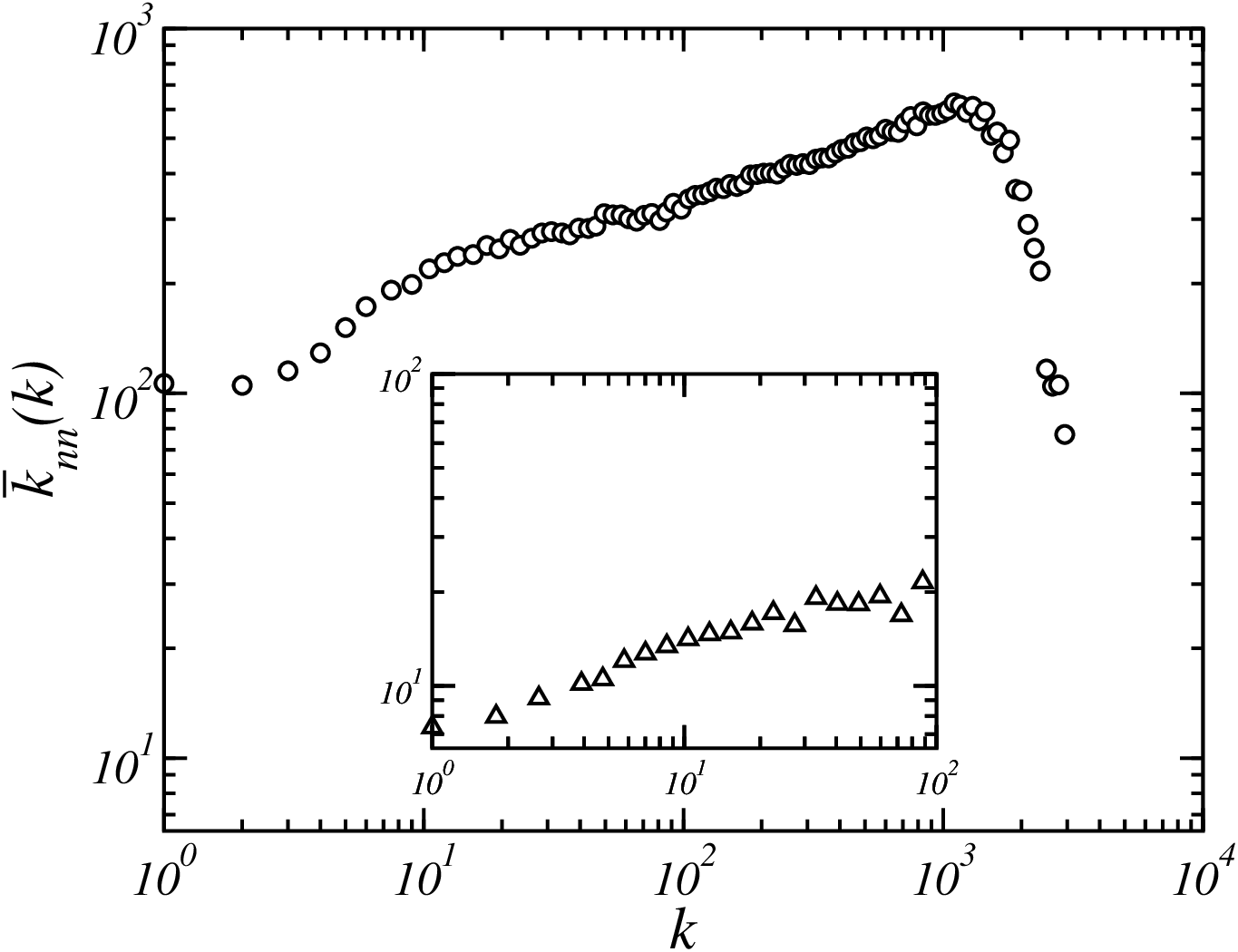}
\caption{Average degree of the nearest neighbors as a function of the
  degree $\bar{k}_{nn}(k)$ for the movie actors collaboration network
  (main plot) and the scientific collaboration network (inset).}
\label{figknn}
\end{figure}

The degree distribution of the one-mode projection of this network,
$P(k)$, is plotted in Fig.~\ref{fpk}.  It has a power law decay with
approximately the same exponent as $Q(q)$, which extends for close to
two decades up to a sharp cutoff at $k_{c} \sim 2000$. The mean degree
of the network is $\langle k \rangle = 78.69$.
The local clustering 
as a function of 
degree is
depicted in Fig.~\ref{figck}. We can observe a flat region, extending
up to degree values close to $10^2$, followed by a rapid decrease.  
The mean clustering of the one-mode projection is $\langle c \rangle = 0.78$ and the clustering coefficient is $c = 0.17$. 
The correlations in
the projected network, 
presented in the form of  
the average degree of the
nearest neighbors of a vertex versus its degree, 
are plotted in Fig.~\ref{figknn}. The
increasing behavior of the function $\bar{k}_{nn}(k)$ is compatible
with the presence of assortative mixing, a fact that is further
confirmed by the value of the Pearson coefficient, $r=0.23$. In
Table~\ref{tab:averages} we summarize the main average values obtained
for this network.


\subsection{Scientific collaboration networks}

The next collaboration network that we analyze is the network of
scientific collaborations collected from the condensed matter preprint
database at Los Alamos \cite{note3}. 
In this
collaboration graph, the actors represent scientists 
which 
have
collaborated in the writing of a scientific paper. The complete
bipartite network is composed by $t = 17828$ papers and $N = 16258$
authors. The distribution of the number of authors in a given paper is
plotted in Fig.~\ref{f1}a). This 
distribution is clearly
exponential, with an average value $\bar{n} = 3.05$. The distribution
of the number of paper written by any given author, Fig.~\ref{f1}b),
shows, on the other hand, an apparent power law behavior, even though
the limited range that it takes (scarcely more than one decade)
precludes the determination of a significant exponent. The average
number of papers written by any author is in this case $\langle q \rangle
=3.35$.

The degree distribution of the one-mode projection of the scientific
collaboration network, plotted in Fig.~\ref{fpk}, shows again a fat-tailed behavior, compatible with a power law. The corresponding
average degree is $\langle k \rangle = 5.85$. The degree-dependent local clustering 
$c(k)$ and the average degree of the nearest neighbors are
explored in Figs.~\ref{figck} and~\ref{figknn}, respectively. This
last result, together with a Pearson $r=0.31$, indicates the presence
of a strong assortative mixing. Additional numerical parameters
characterizing this network are summarized in
Table~\ref{tab:averages}.

\subsection{Board of directorships}

The last collaboration network that we report is the network of
company directors, in which two directors are linked if they sit on
the same board of directors. Table~\ref{tab:averages} reports the data
corresponding to the list of the ``Fortune 1000'' US companies,
obtained from Refs. \cite{newman-review,socialwebs}.  It includes $t =
914$ companies and $N = 7673$ directors. The average number of
directors per company is $\overline{n} = 11.5$.  Both distributions
$Q(q)$ and $P(k)$ can be adjusted by exponential decaying functions,
although the range of values for $q$ and $k$ is quite restricted. The
mean degree of the projected network is $\langle k \rangle = 14.44$. The
clustering coefficient of the one-mode projected network is quite
large, and it shows a clear assortative mixing behavior, as given by a
Pearson correlation coefficient $r=0.28$.

\section{Self-organized collaboration model}\label{s-model}

\subsection{Definition of the model}

To understand the common properties of collaboration networks, we
propose a self-organized growing model.  We exploit two generic
features of collaboration networks: (i) Social collaboration networks
are organized as bipartite graphs.  (ii) Social collaboration networks
are not static entities, but they grow in time by the continuous
addition of new acts of collaboration (movies produced or papers
written), and new actors, that increase the pool of possible
participants in new acts of collaboration.

Using the language of movies to make the description more concrete,
our 
growing bipartite network model 
is defined by the
following rules:
\begin{enumerate}
\item 
At each time step a new movie with $n$ actors is added. 
\item
Of the $n$ actors playing in a new movie, $m$ actors are new, 
       without previous experience. 
\item
The rest $n-m$ actors are chosen from the pool of ``old'' actors 
       with a probability proportional to the number $q$ of movies that 
       they previously starred. 
\end{enumerate} 
The total number of movies is $t$, the ``time''.  The number $n$ may
be either constant or a random variable distributed with a given
distribution $S(n)$.  The number $m$ may also be either constant or a
random variable.  At each time step, the total number of actors
increases as $N \to N + m$.  Thus, the model generates a bipartite
graph of $t$ movie vertices and $N$ actor vertices.  Note that the
proportional preference corresponds to the following practical rule of
selection of actors: A director randomly selects a previous movie and
then chooses at random one of its actors.

\subsection{Analytical results}

One can 
see that the evolution rules of the present model
practically coincide with those of the Simon model \cite{simon55}.
For simplicity, let us assume that the number of actors playing in
each movie is constant and equal to its average value, $n=\bar{n}$, as
well as the number of new actors per movie, $m=\bar{m}$. This
assumption is in fact quite
reasonable 
given the exponential nature of
the $S(n)$ distributions observed empirically. 
We also assume that if the total number of actors is large,  
the probability that two actors selected for a
new movie have already co-starred in other old film is vanishingly
small. Note that, strictly speaking, this 
assumption
is only valid for
uncorrelated networks 
with rapidly decreasing degree distributions.  
Nevertheless, the results obtained with it
provide a good enough approximation to the empirical values (see
Table~\ref{tab:averages}) to justify its introduction.

Within this approximation, since each movie starred by an actor leads
to the acquisition of $\bar{n}-1$ new co-actors, we have a strict
relation between the experience of an actor, $q$, and the total number
of its co-actors (its degree in the projected network), $k$:
\begin{equation}
  k= q \, (\bar{n}-1).
  \label{eq:3}
\end{equation}
In particular, at large $k$ and $q$, when we can
consider both variables 
to be continuous, 
we have
\begin{equation}
  P(k)  \cong \frac{1}{\bar{n}-1}\,Q\left( \frac{k}{\bar{n}-1} \right) \, .
\label{e5}
\end{equation}   

In the limit of large $N$, the total number of edges in the
one-mode projected graph (the number of pairwise coactorships) is $t$
times the number of pairs of actors in a new film, that is, $t \,
\bar{n} \, (\bar{n}-1)/2$, while the total number of actors is $N =
t\, \bar{m}$.  Thus, the mean degree of the one-mode projection
network is
\begin{equation}
  \langle k \rangle  = \frac{\bar{n} (\bar{n}-1)}{\bar{m}} \, .
  \label{e6}
\end{equation} 
Therefore, 
\begin{equation}
  \langle q \rangle = \frac{ \langle k \rangle}{\bar{n}-1} = \frac{\bar{n}}{\bar{m}}.
\end{equation}

As in the Simon model, the connections of this growing
bipartite graph self-organize into a scale-free structure. Quite
similarly to standard derivations for the Simon model, in the large
network limit, the distribution takes the form \cite{simon55,dm03}
\begin{equation}
  Q(q) = (\gamma-1)B(q,\gamma) \, ,  
  \label{e7}
\end{equation} 
where $B(\ ,\ )$ is the $\beta$-function \cite{abramovitz} and
\begin{equation}
  \gamma = 2 + \frac{\bar{m}}{\bar{n}-\bar{m}}.
  \label{eq:2}
\end{equation}
For $q \gg 1$, the asymptotics of $Q(q)$ is
\begin{equation}
Q(q) \sim (q + \gamma-1/2)^{-\gamma} \, , 
\label{e8}
\end{equation} 
so that $\gamma$ is the exponent of the degree distribution.  In the
one-mode projection, this corresponds to
\begin{equation}
P(k) \sim [k + (\gamma-1/2)(\bar{n}-1)]^{-\gamma}\, . 
\label{e9}
\end{equation} 
That is, the projected degree distribution exhibits a power law
behavior, with an off-set $k_0 = (\gamma-1/2)(\bar{n}-1)$. The presence of
this off-set, which may be large for large values of $\bar{n}$, can
hinder the direct evaluation of the exponent $\gamma$.  Therefore, it is
more appropriate to compare the degree distribution with the general
expression Eq.~(\ref{e9}).

To calculate the clustering coefficient, we need to recall our second
assumption: If we consider a particular actor, $i$, who has played in
$q$ movies, in the thermodynamic limit none of his co-actors repeats
twice in different films. This means that in the projected network the
triangles attached to a vertex $i$ can only relate his co-actors
inside each separate movie. The number of such triangles is $q \,
(\bar{n}-1) \, (\bar{n}-2)/2$, while the total number of possible
triangles attached to $i$ is $k \, (k-1)/2$ or, equivalently, $q\,
(\bar{n}-1) \, [q\,(\bar{n}-1)-1]/2$. The local clustering as a
function of the experience of an actor is then given by

\begin{equation}
  c(q) = \frac{\bar{n}-2}{q \, (\bar{n}-1)-1},
  \label{e10}
\end{equation}
which, as a function of $k$, transforms into
\begin{equation}
  c(k) =  \frac{\bar{n}-2}{k-1}.
  \label{e11}
\end{equation}

Then using the definition (6) readily yields the average clustering 

\begin{eqnarray}
    \langle c \rangle   &=&  \sum_{k>1}  P(k) c(k) 
    =  (\bar{n}-2) \, \sum_{k>1} \frac{P(k)}{k-1} \nonumber \\
    &=&
     \frac{\bar{n}-2}{\bar{n}-1} \sum_{q>0} \frac{Q(q)}{q-1/(\bar{n}-1)} .
    \label{e12}
\end{eqnarray} 
On the other hand, to compute the clustering coefficient $c$, defined
in Eq.~(\ref{e3}), we need to estimate the number of triangles
attached to a vertex of degree $k$. As we have seen before, this
number is $q \, (\bar{n}-1) \, (\bar{n}-2)/2$, or, in terms of $k$,
$s(k)= k(\bar{n}-2)/2$. Therefore,
\begin{eqnarray}
c & = & \frac{\sum_q P(q)q(n-1)(n-2)/2}{\sum_q P(q)q(n-1)[q(n-1)-1]/2} 
\nonumber
\\[5pt]
    & = &
\frac{(n-2)\langle q \rangle}{(n-1)\langle q^2 \rangle - \langle q \rangle}
      = 
\frac{(n-2)\langle k \rangle}{\langle k^2 \rangle - \langle k \rangle} 
\, .
\label{e13}
\end{eqnarray}

One can see that the average clustering $ \langle c \rangle$ converges to a
finite value for any degree distribution, since the region of low
degrees makes the main contribution.  Consequently, Eq.~(\ref{e12})
works well even if the degree distribution is fat-tailed. The
clustering coefficient, on the other hand, approaches zero if the
second moment of the degree distribution diverges. This divergence
takes place for $\gamma \leq 3$ in the thermodynamic limit ($N, t \to \infty$).
In this case, $c$ crucially depends on the degree cutoff $k_c$ (or
$q_c$) in the form $c \sim k_c^{-(3-\gamma)}$.  Note that formula
(\ref{e13}) may underestimate the value of the clustering coefficient
if the degree distribution is fat-tailed and $k_c$ is large.

\section{Results and comparison with real networks}\label{s-results}

To check the validity of our model, we proceed to compare the
empirical data on collaboration networks with the predictions made in
the previous Section, as well as with numerical simulations of
the model.  The analytic predictions are specified in terms of two
parameters, the average number of actors per collaboration act
$\bar{n}$ and the average number of new actors $\bar{m}$. If all these
actors are recruited at a constant rate, then we have in average
$\bar{m} = N/t$ new actors per collaboration act.  From these two
parameters, using the results of the previous Section, we can compute
our predictions for all the properties of the networks described in
Section~\ref{sec:empir-data-coll}. When performing numerical
simulations of the model, and in order to avoid discreteness, we use
randomly distributed $m$ and $n$. Their distributions are taken to be
exponential with averages $\bar{m}$ and $\bar{n}$ respectively. This
functional form corresponds to that of the distribution $S(n)$
empirically observed for actor and scientific collaboration networks
(see Fig.~\ref{f1}a). For the company directorship network on the
other hand, we do not count with an empirical form of $S(n)$. Hence,
we checked both exponential and Poisson distributions. The global
characteristics of the networks generated with this last distribution
suit better their empirical counterparts. The results of the
comparison between empirical data, theoretical predictions and
simulations are summarized in Table~\ref{tab:averages}.

\begin{figure}
\epsfxsize=80mm
\epsffile{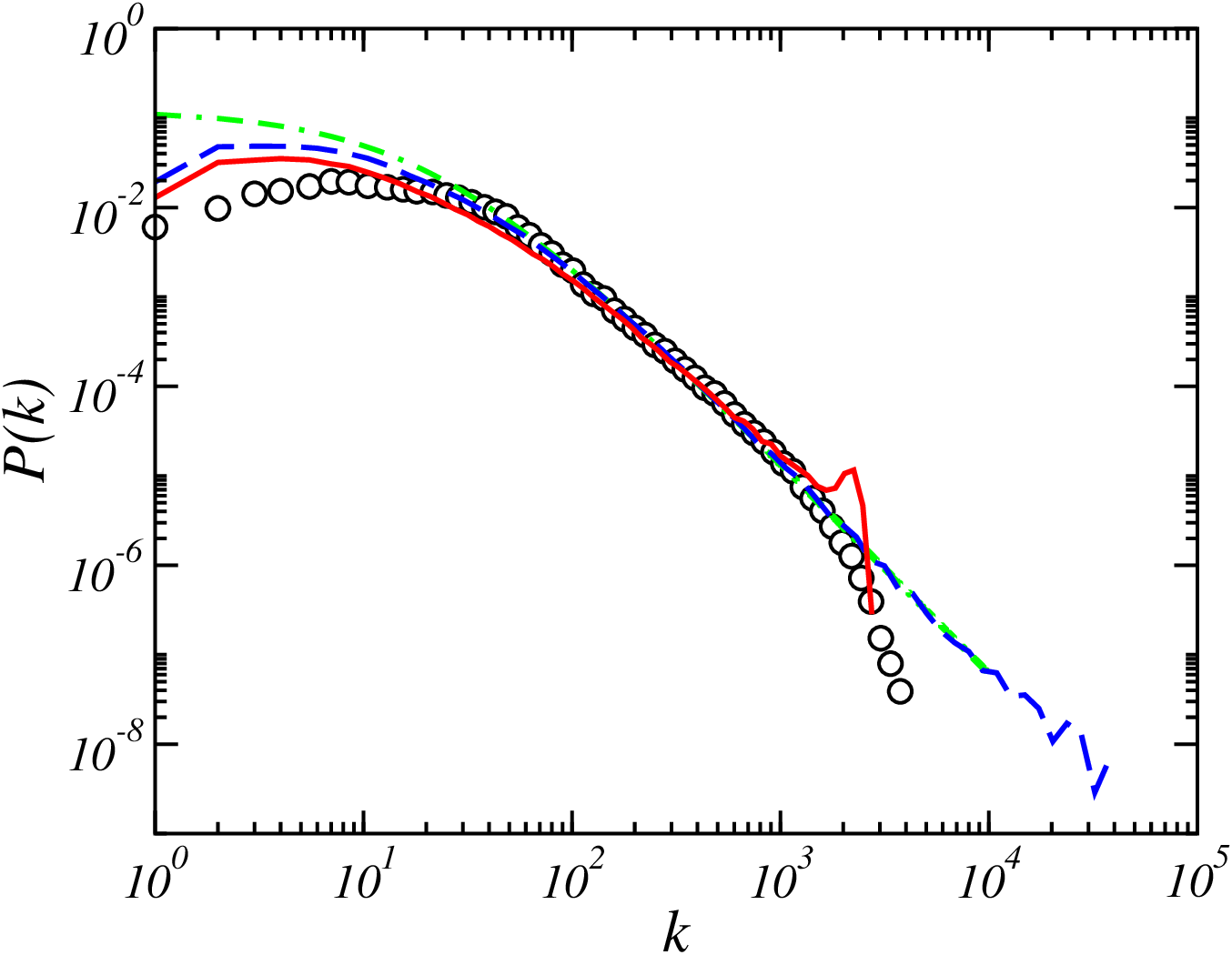}
\caption{Comparison among the degree distribution $P(k)$ for the
  empirical movie actors collaboration network (circles), for the
  theoretical prediction of section IV (dot--dashed
  line), and for simulations of the original model (dashed curve) and 
  the version with aging (solid line).}
\label{f3}
\end{figure}

\begin{table*}
\caption{
\label{tab:averages}Comparison of calculations with empirical data for
the movie, scientific collaboration, and  codirectorship networks,
and the simulations of the model.} 
\begin{ruledtabular}
\begin{tabular}{cccccccccccc}
&
movie actors&
analytic&
numeric&
numeric&
coauthors&
analytic&
numeric&
numeric&
directors&
analytic&
numeric\\
&
network&
results&
results&
with aging&
network&
results&
results&
with aging&
network&
results&
results\\
\hline 
$t$&
$118477$&
& 
$10^5$& 
$10^5$& 
$17828$&
&
$10^5$&
$10^5$&
$914$&
&
$10^5$\\
$N$&
$382219$&
&
&
&
$16258$&
&
&
&
$7673$&
&
\\
$\bar{{n}}$&
$12.33$&
$12.33$&
$12.33$&
$12.33$&
$3.05$&
$3.05$&
$3.05$&
$3.05$&
$11.5$&
$11.5$&
$11.5$
\\
$\bar{{m}}$&
&
$3.23$&
$3.23$&
$3.23$&
&
$0.91$&
$0.91$&
$0.91$&
&
$8.39$&
$8.39$
\\
$q_{c}$&
$\sim 10^{2}$&
&
&
$10^{2}$&
$\sim 15$&
&
&
$15$&
$\sim 10$&
&
\\
$\langle q\rangle $&
$3.82$&
$3.82$&
$4.44$&
$4.39$&
$3.35$&
$3.35$&
$3.70$&
$3.69$&
$ $&
$1.37$&
$1.48$
\\
$\langle k\rangle $&
$78.69$&
$43.25$&
$75.05$&
$85.67$&
$5.85$&
$6.87$&
$8.45$&
$8.93$&
$14.44$&
$14.39$&
$17.10$\\
$\gamma $&
&
$2.35$&
&
&
&
$2.43$&
&
&
&
$4.70$&
\\
$\langle c\rangle $&
$0.78$&
$0.71$&
$0.76$&
$0.70$&
$0.64$&
$0.68$&
$0.50$&
$0.43$&
$0.88$&
$0.87$&
$0.86$\\
$c$&
$0.17$&
$0.06$\footnote{The cutoff $k_{c}\sim 10^{3}$ was used.}&
$0.037$&
$0.08$&
$0.36$&
$0.08$\footnote{The cutoff $k_{c}\sim 10^{2}$ was used.}&
$0.026$&
$0.09$&
$0.59$&
$0.5$
&
$0.32$\\
$r$&
$0.23$&
&
$-0.13$&
$0.14$&
$0.31$&
&
$-0.08$&
$0.40$&
$0.28$&
&
$0.11$\\
\end{tabular}
\end{ruledtabular}
\end{table*}

We observe an agreement between the model predictions and the
empirical results for the mean clustering $\langle c\rangle$.  Note some
deviations in the mean degree $\langle k\rangle$ of one-mode projected networks.
These discrepancies are due to the fact that in our analysis we
neglect the probability that some actors for a new film have
previously co-starred in the same movies.  For the net of
codirectorships, the computed clustering coefficient $c$ is in a
reasonable agreement with the empirical value.  In the other networks,
however, the calculated values of $c$ are severely underestimated. The
reason for this is the poor quality of the approximate formula
(\ref{e13}) for this model in the case of a fat-tailed degree
distribution (see discussion in Sec.~\ref{s-model}).

The exponents of the projected degree distribution are $\gamma=2.35$,
$\gamma=2.43$ and $\gamma=4.7$ for the movie, coauthorship and codirectorship
networks, respectively.  The exponent larger than $3$ in this last
case is compatible with the exponential decay observed empirically.
The range of the empirical degree distribution in the
coauthorship network is too small to compare with the asymptotic
expression Eq.~(\ref{e9}).  So, we make this comparison only in the
case of the movie actors network, Fig.~\ref{f3}.  As can be seen, the
agreement between the theoretical and empirical distributions is
notorious.  Only at very small and very large values of the degree a
certain discrepancy can be noticed essentially due to the continuous
degree approximation employed and the presence of a cutoff in the
empirical distribution, respectively.  This upper cutoff is inevitable
due to two factors: (i) an actor physically cannot have an infinite
number of co-stars, and (ii) finite size effects restrict the degrees
of vertices (see, e.g., Ref.~\cite{mariancutofss}).

\begin{figure}[b]
\epsfxsize=80mm
\epsffile{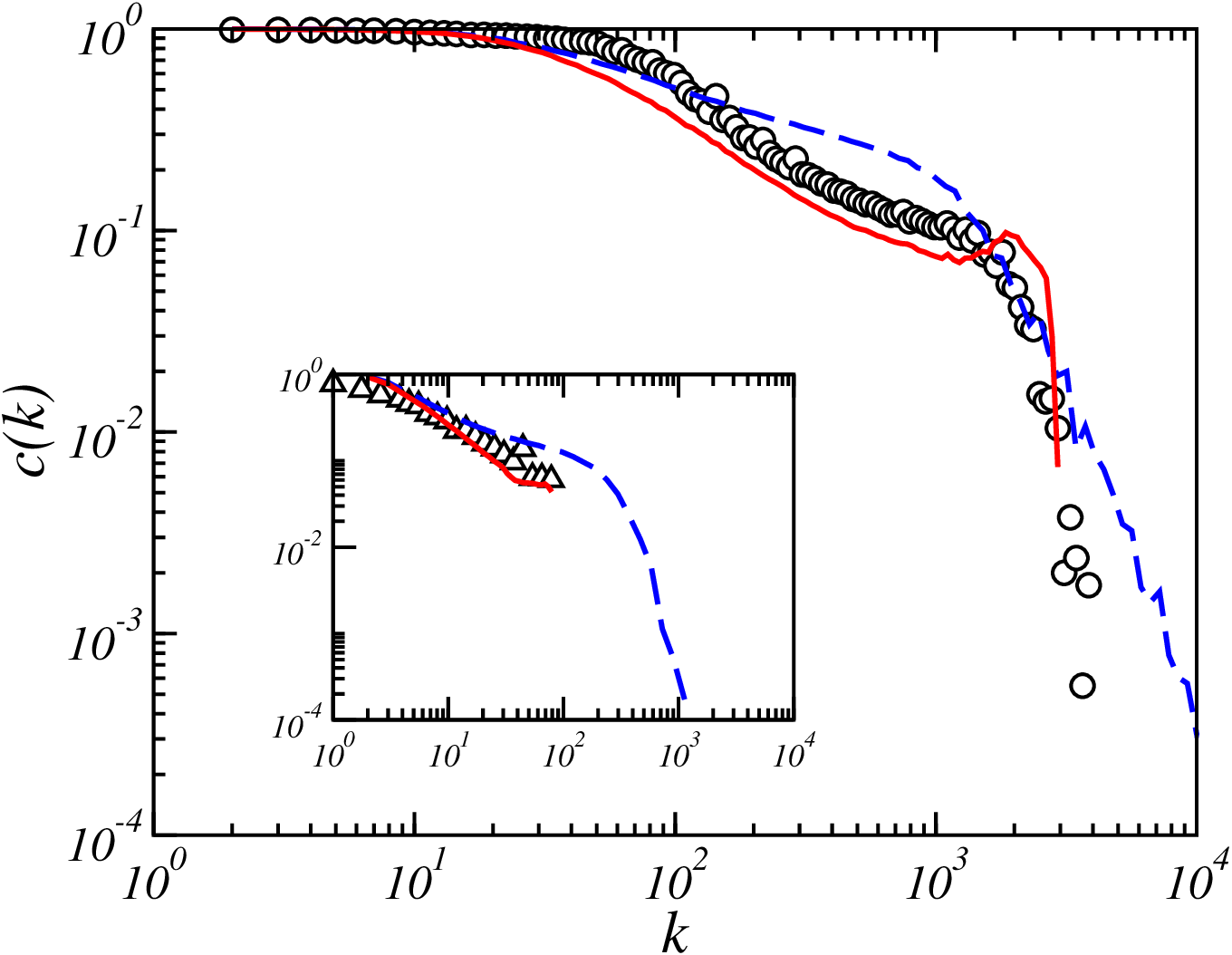}
\caption{Comparison among the clustering coefficient as a function of
  the degree $c(k)$ for the empirical movie actors collaboration
  network (circles), for the simulations of the original version of
  the model (dashed line), and for the model with aging (solid
  line). The main plot is for the actor co--starring network and the inset for 
  the scientific collaboration network.}
\label{f5}
\end{figure}

\begin{figure}
\epsfxsize=85mm
\epsffile{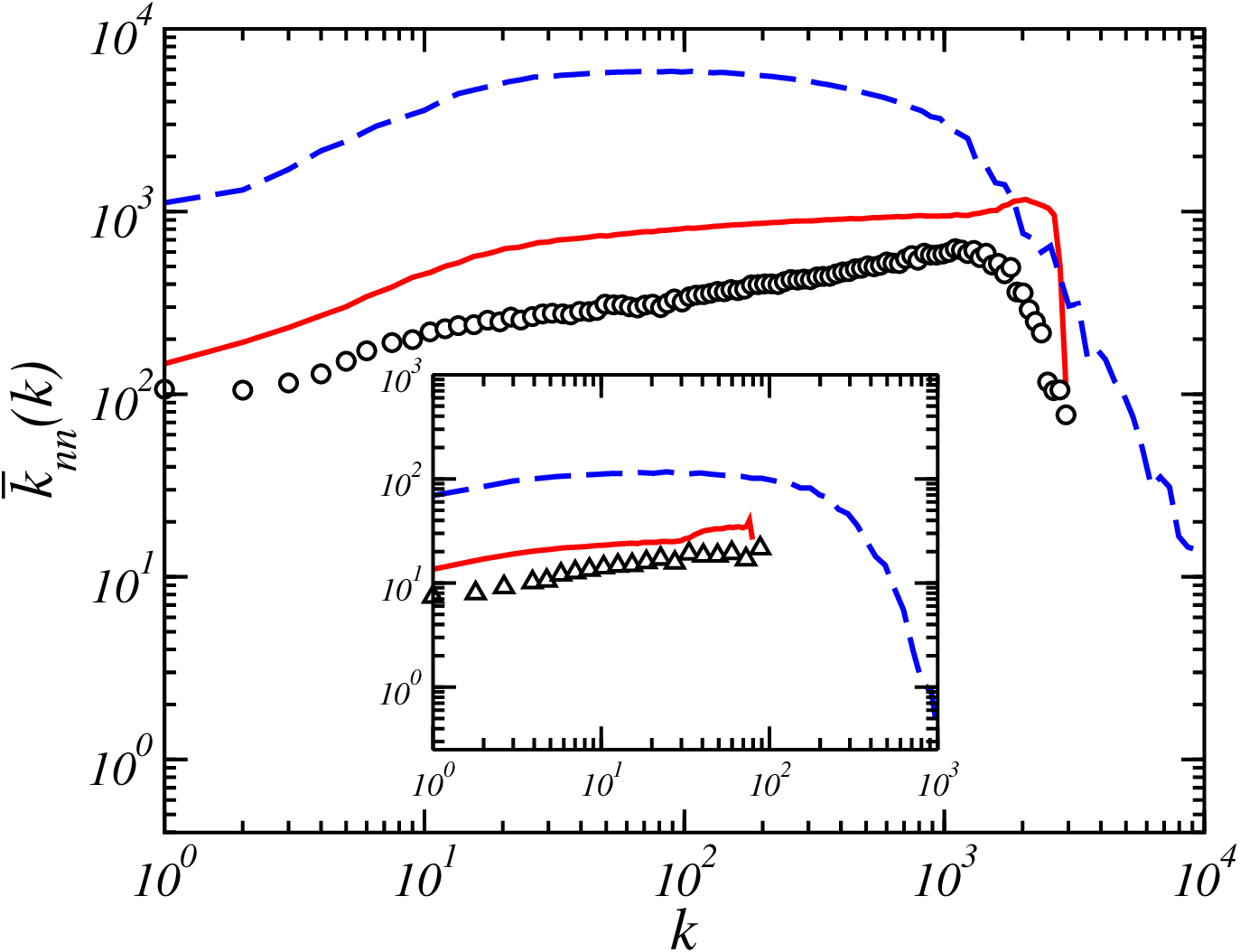}
\caption{Comparison among the  average degree of the nearest neighbors
  as a function of the degree $\bar{k}_{nn}(k)$ for the empirical
  movie actors collaboration network (circles), the simulations of the 
  original version of the model (dashed line), and for
  the model with aging (solid line). The main plot is for the actor co--starring network and the inset for 
  the scientific collaboration network.}
\label{f6}
\end{figure}

The empirical local clustering and its analogue obtained by
numerical simulations (Fig.~\ref{f5}) demonstrate a more complex
dependence on degree that the simple estimate of Eq. (\ref{e11}).
From Fig.~\ref{f5} we see that the function $c(k)$ computed from the
model follows a slower decay than the corresponding empirical function
for the movie actors network.  The results for the mean degree of the
nearest neighbors of a vertex as a function of its degree,
$\bar{k}_{nn}(k)$, are represented in Fig.~\ref{f6}.  Unexpectedly,
apart from a small region for very small values of $k$,
$\bar{k}_{nn}(k)$ decreases with degree, and the Pearson correlation
coefficient is negative (see Table~\ref{tab:averages}).  So that,
unlike real-world collaboration graphs, networks with a fat-tailed
degree distribution, generated by the simplest version of our model,
show disassortative mixing.  On the contrary, in the case of
directorship networks, the model provides positive values for the
Pearson coefficient $r$, in agreement with the empirical results,
though a little lower (see discussion in the next Section).

\section{Self-organized model with aging}\label{s-model-age}

At least in one aspect, 
the model presented above is a serious oversimplification of the mechanism
underlying the growth of social collaboration networks. It allows an
analytical treatment but leads to 
problems in the comparison with
the 
clustering coefficient 
and, 
especially, degree--degree  
correlations of empirical networks. The most
important missing point is probably the aging of individual agents. This
issue is evident in the case of the movie actors network (although it can
be observed for the scientific collaboration network too). The Internet
Movie Database site, from which the actor collaboration network was
extracted, contains information spanning the whole century of the
history of cinema, from Louis Lumiere to the most recent Hollywood
productions.  Considering that actors have a finite professional life
time, it is unrealistic to allow them to take part in a movie
irrespective of their age.

If we take into account this fact, two main
consequences are immediately expected. On the one side, there should be
an upper cutoff in the $Q(q)$ distribution corresponding to the
professional life expectation of actors, as actually it is found in empirical
distributions, and, in addition, not all
actors may work together: only those who are contemporaneous. 
Obviously,
this phenomenon affects much less the codirectorship network because of 
its exponential degree distribution.

To introduce this new ingredient in the model, we must first assume an
aging rate for individual agents. The most straightforward way to do
so is to suppose that the time is directly equivalent to the
experience $q$. Actually, in more realistic situations, it may happen
that each agent has its own aging rhythm.  However, the latter version
of aging would make the model more complex.  Once time is identified,
we must consider a survival probability distribution for agents. In
parallel with biological systems, we will assume an almost sure
survival till a certain age $Q_0$ and an exponential decay hereafter.
The modification of the model then requires of two new parameters: the
cutoff $Q_0$ and the characteristic time of the exponential decay
$\tau$. The rest of the model remains the same. That is, in each step a
new movie is produced, $m$ actors are new and the rest of them $n-m$
are chosen at random with a probability proportional to their
experience.  In addition, we assume that the actors become inactive,
i.e. they cannot be chosen again for new movies, with a probability
given by the complementary of the survival distribution for their 
particular age $q$.

We carried out simulations with the new version of the model.  $Q_0$
was fixed at $100$ for the actors network and at $15$ for scientific
collaborations to agree with the cutoff observed in the empirical
distributions $Q(q)$ of these networks. The value of the other
parameter, $\tau$, is not so easy to establish from phenomenological
data, therefore we check several characteristic times. For the sake of
concreteness, let us focus on the results obtained with $\tau = 50$ for
actor co-starring and with $\tau = 7$ for scientific coauthorships,
which are realistic values compatible with the final decay of the
$Q(q)$ empirical distributions. Actually, using $\tau$ two times bigger
we did not observe essential differences in the properties of the
networks.  Moreover, a simple exponential survival probability (i.e.,
with the only parameter $\tau$) also provides similar approximate values
of the clustering coefficient and the Pearson coefficient.  However,
it does not allow to satisfactory describe the whole degree
distribution. Note that our choice of the aging parameters $Q_0$ and
$\tau$ does not actually mean fitting of our final results, which are
the clustering and degree--degree correlation characteristics.
Indeed, the values of $Q_0$ and $\tau$ were chosen only to properly
describe the degree distribution in the range of large degrees.

In Fig.~\ref{f5}, the local clustering is plotted as a function of the
degree for a network with aging and for the empirical actor network.
The dependence $c(k)$ adjusts better to empirical data, and the
computed clustering coefficients are closer to the empirical ones (see
Table~\ref{tab:averages}).  The improvement on these coefficients is
understandable.  As one can see from our simple analytical
estimations, the direct introduction of the cutoff in the degree
distribution seriously improves the values of the clustering
coefficients.

\begin{figure}
\epsfxsize=80mm
\epsffile{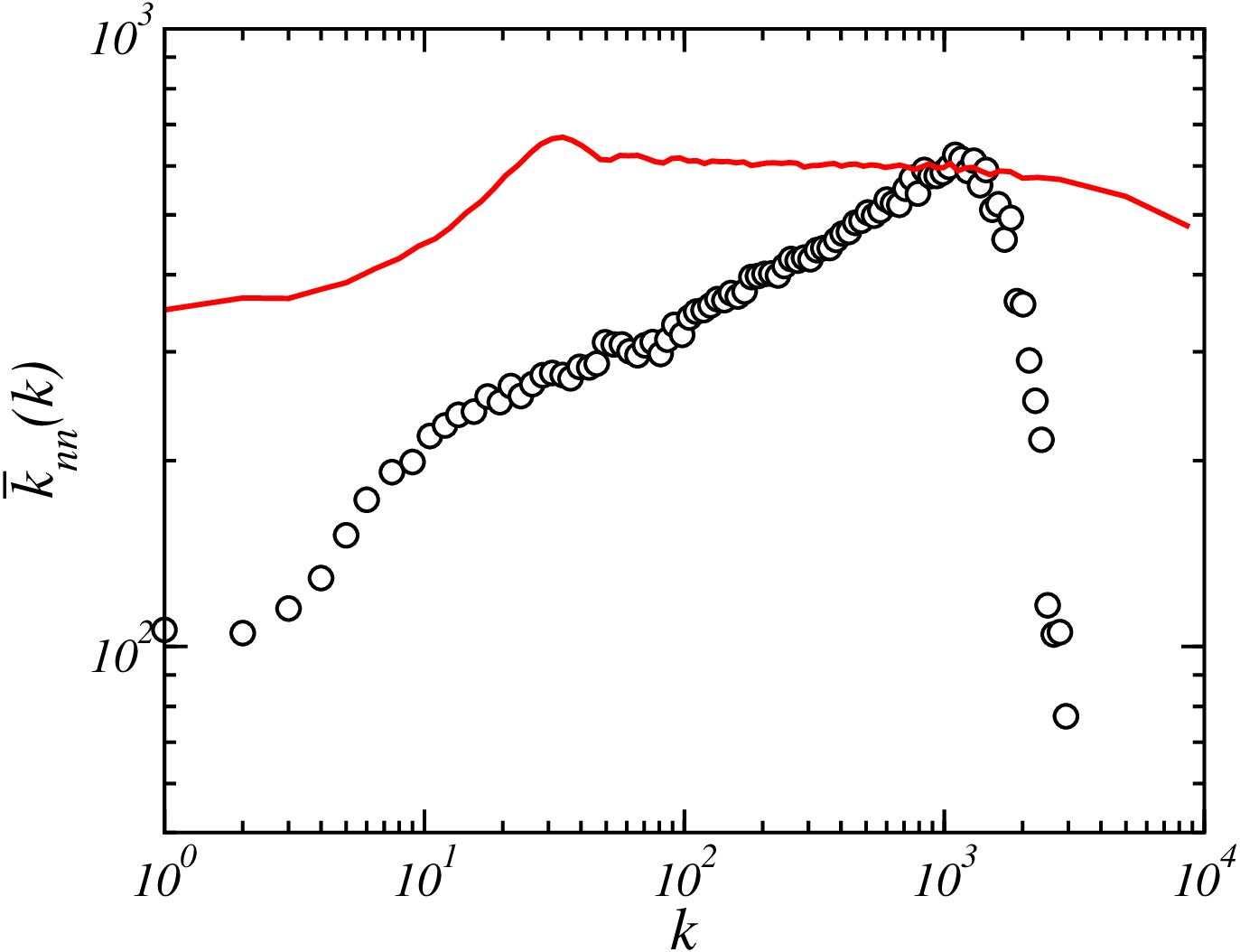}
\caption{Average degree of the nearest neighbors as a function of the
  degree $\bar{k}_{nn}(k)$ for an uncorrelated bipartite graph with
  the same degree distributions for both types of vertices as
  generated by our model (solid line). For comparison, the same quantity is
  displayed for the empirical 
  actor network (circles).}
\label{uncorrelated}
\end{figure}

A far more important point is that the aging changes the type of
degree--degree correlations.  In the version of the model with aging,
the computed dependence of the mean degree of the nearest neighbor of
a vertex on its degree properly describes the empirical dependence, as
may be seen in Fig.~\ref{f6}.  As a result, the computed values of the
Pearson correlation coefficients turns out to be positive (assortative
mixing) and close enough to the empirical values (see
Table~\ref{tab:averages}).  One should note that in the framework of
the configuration model of an uncorrelated bipartite network, this
agreement is impossible.  We have checked this claim in the following
way: We have measured the degree--degree correlations in the one-mode
projection resulting from an uncorrelated bipartite graph with the
same degree distributions for both types of vertices as generated by
our model.  In contrast to the self-organized model, see
Fig.~\ref{uncorrelated}, the curve ${\bar k}_{nn}(k)$ turns out to be
nearly flat, the Pearson coefficient being close to zero.
This signals that the degree--degree
correlations are practically absent in this case.

\section{Conclusion}\label{s-conclusions} 

In summary, we have studied a minimal model of evolving,
self-organizing collaboration networks. This model is not based on a
static perspective as was the configuration model, but on a dynamical
mechanism to construct the network. Besides, its basic constituents are
preferential attachment and the bipartite structure of social 
networks. Our results show that the
self-organized model offers a good starting point to explain
existing empirical data. The model was compared with empirical results
for a number of real networks, namely a network of scientific
coauthorships, a network of movie actor collaborations and a network
of company codirectorships.

We have shown that, apart of a generic bipartite structure and the
growth factor, one more element has to be taken into account in order
to explain the empirical observations on the clustering and
degree--degree correlations in collaboration networks.  This key
factor is the aging of collaborators.  We demonstrate that in
collaboration networks this effect is responsible for the positive
(assortative) degree-degree correlations.  We conclude that
assortative mixing, which is generally observed in collaboration
networks, is produced by the combination of their bipartite structure
and the aging of the collaborators.

One should note that, in principle, even uncorrelated bipartite graphs
(the configuration model) have correlated one-mode projections.
However, the specific degree--degree correlations in these projections
are quite weak. In other words, the configuration model graphs with
degree distributions typical for movie actor nets show neither
assortative nor disassortative mixing (they have $r \approx 0$).  In
contrast, our self-organized model provides correlated bipartite
graphs, which, under natural assumptions, have one-mode projections
with realistic structure and realistic correlations.

\begin{acknowledgments}
  We wish to thank M. E. J. Newman for making available the data for
  the cond-mat collaboration network, and M. Bogu\~n\'a and A. Vespignani
  for helpful comments and discussions.  Support from the Portuguese
  Research Council under grant SFRH/BPD/14394/2003 and the European
  Commission - Fet Open project COSIN IST-2001-33555 is acknowledged.
  R.P.-S. acknowledges financial support from the Ministerio de
  Ciencia y Tecnolog{\'\i}a (Spain), and from the DURSI, Generalitat de
  Catalunya (Spain).
\end{acknowledgments}


\begin{thebibliography}{10}
\bibitem{note1}
If various
kinds of collaborations are present, the collaboration network may
be represented by a multi-partite graph. Here we restrict
ourselves to the simplest situation.
\bibitem{note4} Note,
however, that $\bar{k}_{nn}(k)$ may be non-monotonous.
\end{thebibliography}
\end{document}